\documentclass[review]{elsarticle}
\usepackage{lineno,hyperref}
\modulolinenumbers[5]

\usepackage[version=3]{mhchem} 
\usepackage{subfigure,xfrac,float}

\usepackage{numcompress}

\begin{document}

\begin{frontmatter}

\title{Structural studies of mesoporous \ce{ZrO2}-\ce{CeO2} and \ce{ZrO2}-\ce{CeO2}/\ce{SiO2} mixed oxides for catalytic applications}

\author[mymainaddress]{R. Bacani\corref{mycorrespondingauthor}}
\address[mymainaddress]{Departamento de F\'isica Aplicada, Instituto de F\'isica, Universidade de S\~ao Paulo, Travessa R da Rua do Mat\~ao, No. 187, Cidade Universit\'aria, 05508-090, S\~ao Paulo, Brazil.}
\cortext[mycorrespondingauthor]{Corresponding author}
\ead{rbacani@if.usp.br}



\author[mysecondaryaddressa]{T. S. Martins}
\address[mysecondaryaddressa]{Departamento de Ci\^encias Exatas e da Terra, Instituto de Ci\^encias Ambientais, Qu\'imicas e Farmac\^euticas, Universidade Federal de S\~ao Paulo - UNIFESP, Rua S\~ao Nicolau, 210, 2o andar, Diadema, 09913-030, S\~ao Paulo, Brazil.}

\author[mymainaddress]{M. C. A. Fantini}

\author[mysecondaryaddressb]{D. G. Lamas}
\address[mysecondaryaddressb]{CONICET / Escuela de Ciencia y Tecnolog\'ia, Universidad Nacional de General San Martín. Mart\'in de Irigoyen 3100, Edificio Tornav\'ia, Campus Miguelete, (1650) San Mart\'in, Pcia. de Buenos Aires, Argentina}

\begin{abstract}
In this work the synthesis of \ce{ZrO2}-\ce{CeO2} and \ce{ZrO2}-\ce{CeO2}/\ce{SiO2} were developed, based on the process to form ordered mesoporous materials such as SBA-15 silica. The triblock copolymer Pluronic P-123 was used as template, aiming to obtain crystalline single phase walls and larger specific surface area, for future applications in catalysis. Small angle X-ray scattering and X-ray diffraction results showed a relationship between ordered pores and the material crystallization. 90\% of \ce{CeO2} leaded to single phase homogeneous ceria-zirconia solid solution of cubic fluorite structure (Fm$\bar{3}$m). The \ce{SiO2} addition improved structural and textural properties as well as the reduction behaviour at lower temperatures, investigated by X-ray Absorption near edge spectroscopy measurements under \ce{H2} atmosphere.
\end{abstract}


\end{frontmatter}


\section{Introduction}

Mesoporous materials have interesting properties, such as high surface area, adjustable size and shape of pores, several different structures and compositions which provide potential applications in catalysis, adsorption, sensors, drug delivery and nano devices\cite{schuth2014}. The extension of silicon-based materials synthesis to other metal oxides is extensively pursued, since ordered mesoporous materials such as MCM-41 and SBA-15 present excellent textural/structural properties\cite{schuth2014,kresgue,vartuli,sayari1997,ciesla1999,schuth2005}. 

Zirconia-Doped Ceria (ZrO$_2$-x\%CeO$_2$, ZDC) is a largely studied solid solution due to its physical-chemical properties that are useful for applications in three-way catalysts and as anodes in Solid Oxide Fuel Cells (SOFC). Among these properties, a high surface area is pursued since the gas permeation in the material allows a high gas reaction rate. Also, ceria-based materials present the ability to store/release oxygen during rich/poor fuel conditions, which is correlated to the Ce$^{4+} \leftrightarrow$ Ce$^{3+}$ redox reaction\cite{fornasiero1996,fornasiero1998,kaspar2000,kaspar2005,lamas2010,fornasiero2014}.

Several authors reported the synthesis of mesoporous ceria-zirconia materials\cite{stucky1999, chen2006,yuan2007,park2008,yuan2009,li2009} but few presented a detailed study of the porous network via SAXS and the evolution of the pore ordering, regarding the materials crystallization. Therefore, the aim of this work was to explore different synthesis methods to produce ZDC samples and fully characterize their physical chemical properties for future applications. In particular, the strategy to build a \ce{SiO2} palisade was analysed as an attempt to avoid shrinkage during calcination.

\section{Experimental} \label{sec:S}

\subsection{\ce{ZrO2}-\ce{CeO2} and \ce{ZrO2}-\ce{CeO2}/\ce{SiO2} preparation}

\paragraph{Anhydrous chloride method:}2 g of Pluronic P-123 (BASF) was previously stirred in 10 mL of ethanol, a total of 13.8 mmol of anhydrous \ce{ZrCl4} and \ce{CeCl3} (Aldrich) were solubilized in another 10 mL of ethanol and added to the polymer mixture (Ce/Zr molar ratios of 0.5 and 0.9). Approximately 2 mL of water were added to the solution, and the Zr/Ce precursor salts and polymer were stirred for 24 hours. The solution was placed in an open Petri dish for thermal treatment (with water vapour) in the oven at 60 $^{\circ}$C, until dried. The samples were identified as Z50AC and Z90AC (50 and 90 mol\% \ce{CeO2}).

\paragraph{Hybrid method (10 mol\% Si):} 2 g of Pluronic P-123 was previously stirred in 10 mL of ethanol and 1 mL of \ce{HCl} (2 mol/L). Then 0.15 mL (total 10 mol\%) of TEOS (Tetraethyl orthosilicate, Aldrich), the Si precursor, was added drop wise in the solution and stirred together for 30 minutes. After that 6 mmol of \ce{ZrCl4} and \ce{CeCl3}$\cdot$7\ce{H2O} (Aldrich) were solubilized in another 10 mL of ethanol and added to the polymer mixture (Ce/Zr molar ratio of 0.9). \ce{NH4OH} was added to achieve a pH of 3 and the remaining solution was stirred for 24 hours. The solution was placed in a Teflon autoclave for hydrothermal treatment in the oven at 60 $^{\circ}$C for 1 day, and dried in a water bath for 5 hours. The sample was identified as Z90H10.

\paragraph{Palisade method (10 and 30 mol\% Si):} based on the last synthesis, the P-123 polymer and TEOS (total 10 and 30 mol\%) were stirred previously for 6 hours before adding the same Zr/Ce salt precursors with Ce/Zr molar ratio of 0.9. Then \ce{NH4OH} was added to achieve a pH of 3 and the remaining solution was stirred for 24 hours. The solution was placed in a Teflon autoclave for hydrothermal treatment in the oven at 60 $^{\circ}$C for 1 day, and dried in a water bath for 5 hours. The samples were identified as Z90P10 and Z90P30.

The calcination process to remove all the samples' template was performed in a tubular oven, with a temperature heating rate of 1 $^{\circ}$C/min, until 540 $^{\circ}$C in \ce{N2} atmosphere, an isotherm in 540 $^{\circ}$C with 2 hours in \ce{N2} and 2 hours in air.

\subsection{\ce{ZrO2}-\ce{CeO2} characterization}

In-situ The Small angle X-ray scattering (SAXS) experiments were carried out at the SAXS-1 beamline of the Brazilian Synchrotron Light facility, LNLS. The measurements were performed with Si-(111) crystal as a monochromator, choosing $\lambda$=1.608 \AA. The sample to detector distance was 800 mm and the detector was a bi-dimensional CCD camera. The oven was programmed for a heating rate of 1 $^{\circ}$C/min, with an isotherm at 550 $^{\circ}$C. The sample was set under vacuum ($\sim$10$^{-3}$ mbar). The data were normalized by the attenuation factor and the total intensity.
	
The X-ray powder diffraction (XRD) measurements were carried out at the XPD beamline of the LNLS. The wavelength was set at 1.5498 \AA. High-intensity (low-resolution) configuration, without crystal analyser was chosen with 2$\theta$ from 18$^{\circ}$ to 102$^{\circ}$, with a 0.05$^{\circ}$ step, and counting times between 2.5-3.0 s/step.

\subsection{\ce{ZrO2}-\ce{CeO2}/\ce{SiO2} characterization}

The SAXS measurements were carried out using a NANOSTAR (Bruker) sealed Cu tube ($\lambda$=1.5418 \AA) operating at 40 kV and 30 mA, with a multi-filament Hi-STAR bidimensional detector. The point focus geometry was used; the system was collimated by 3 pinholes and a cross-coupled Goebel-mirror system. The set up holds a vacuum path between the sample chamber and the detector. The sample to detector distance was 650 mm, therefore with scattering vector (\textbf{q}) values ranging from 0.012 \AA$^{-1}$ to 0.35 \AA$^{-1}$. All the data were normalized by the measuring time and corrected for absorption effects.

The XRD measurements were performed at an Ultima Plus (Rigaku) equipment with conventional copper tube ($\lambda$=1.5418 \AA), operating at 40 kV and 30 mA, with 2$\theta$ from 10$^{\circ}$ to 100$^{\circ}$ with a 0.05$^{\circ}$ step, and counting time of 5 s/step. Rietveld's powder structure refinement analysis was performedfor the data of samples calcined until 540 $^{\circ}$C. The software Fullprof\cite{fp} was used to refine the structural parameters through a least-squares method. The peak shape was assumed to be a pseudo-Voigt function with asymmetry parameters. The background of each pattern was fitted by a 5 degree polynomial function. Isotropic atomic displacement (temperature) factors were assumed. The least-square method was adopted to minimize the difference between the observed and simulated powder diffraction pattern. The R's values indicated the agreement between the observed and calculated quantities. The refinement was done, reducing R's values and the quality factor goodness of fit (S$_{GoF}$), until they reach the convergence\cite{rietveld3,rietveld4,rietveld6,rietveld7,rietveld8}. 

Nitrogen adsorption isotherms were obtained for all samples (with and without Si) with a NOVA (Quantachrome) porosimeter. Thermal treatment to dry the samples was performed for 2 hours at 200 $^{\circ}$C and the measurements were taken in 77 K (\ce{N2}). The pore size distribution, pore volume and pore radius were calculated using the BJH method\cite{bjh}. The specific surface area was calculated using the BET method\cite{bet,sing}.

The cerium oxidation state was evaluated by in situ XANES experiments at the Ce L$_{III}$-edge. The spectra were collected at the D04B- XAFS1 beamline of LNLS in transmission mode using a Si(111) monochromator. The samples with \ce{SiO2} were diluted with boron nitride (BN), and pressed into 15 mm diameter pellets (around 4-5 mg of sample and 70 mg of BN). The pellets were placed in a quartz tubular furnace sealed with Kapton windows. These measurements were acquired during a temperature-programmed reduction (TPR) reaction under a 5\% \ce{H2}/He gas mixture (total flow of 50 mL/min) at temperatures in the range of 25 to 500 $^{\circ}$C at a heating rate of 10 $^{\circ}$C/min. The data were analysed with the WinXAS software\cite{winxas}.

\section{Results and discussion}

\subsection{\ce{ZrO2}-\ce{CeO2}}

\begin{figure}[H]
\centering
  \includegraphics[width=0.7\textwidth]{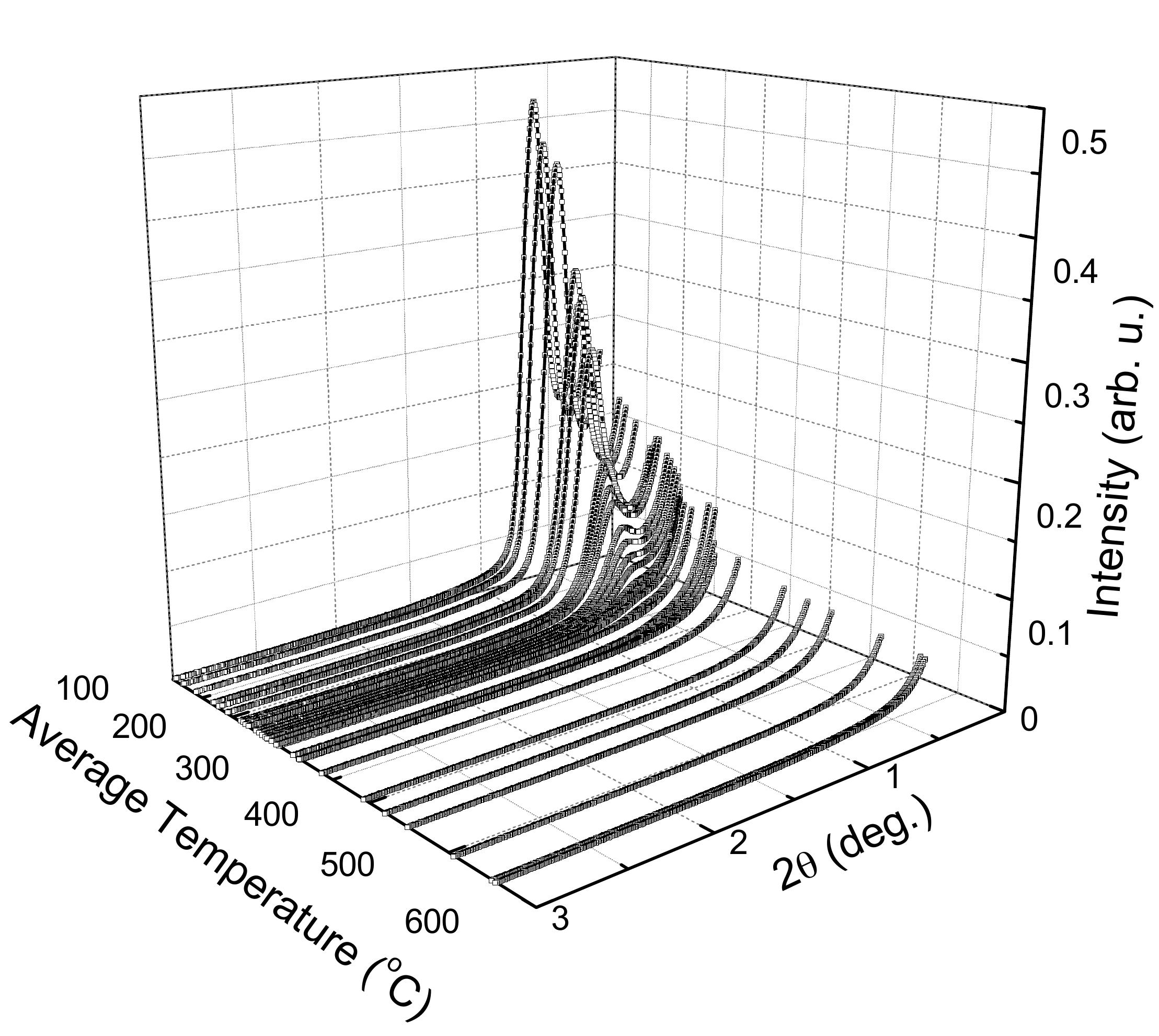}
  \caption{SAXS curves during the calcination process for the Z50AC sample, with 50\% \ce{CeO2}.}
  \label{fgr:zr50s}
\end{figure}

In-situ SAXS experiments were carried out in order to analyse the process of template removal from the samples. Both as-synthesized samples present one diffraction peak, revealing partial spatial correlation between its pores before the calcination process. For the 50\% \ce{CeO2} content (sample Z50AC) this peak decrease as the temperature increases, as shown in figure \ref{fgr:zr50s}. The partially ordered mesopores structure was maintained up to 300 $^{\circ}$C.

\begin{figure}[H]
\centering
  \includegraphics[width=0.7\textwidth]{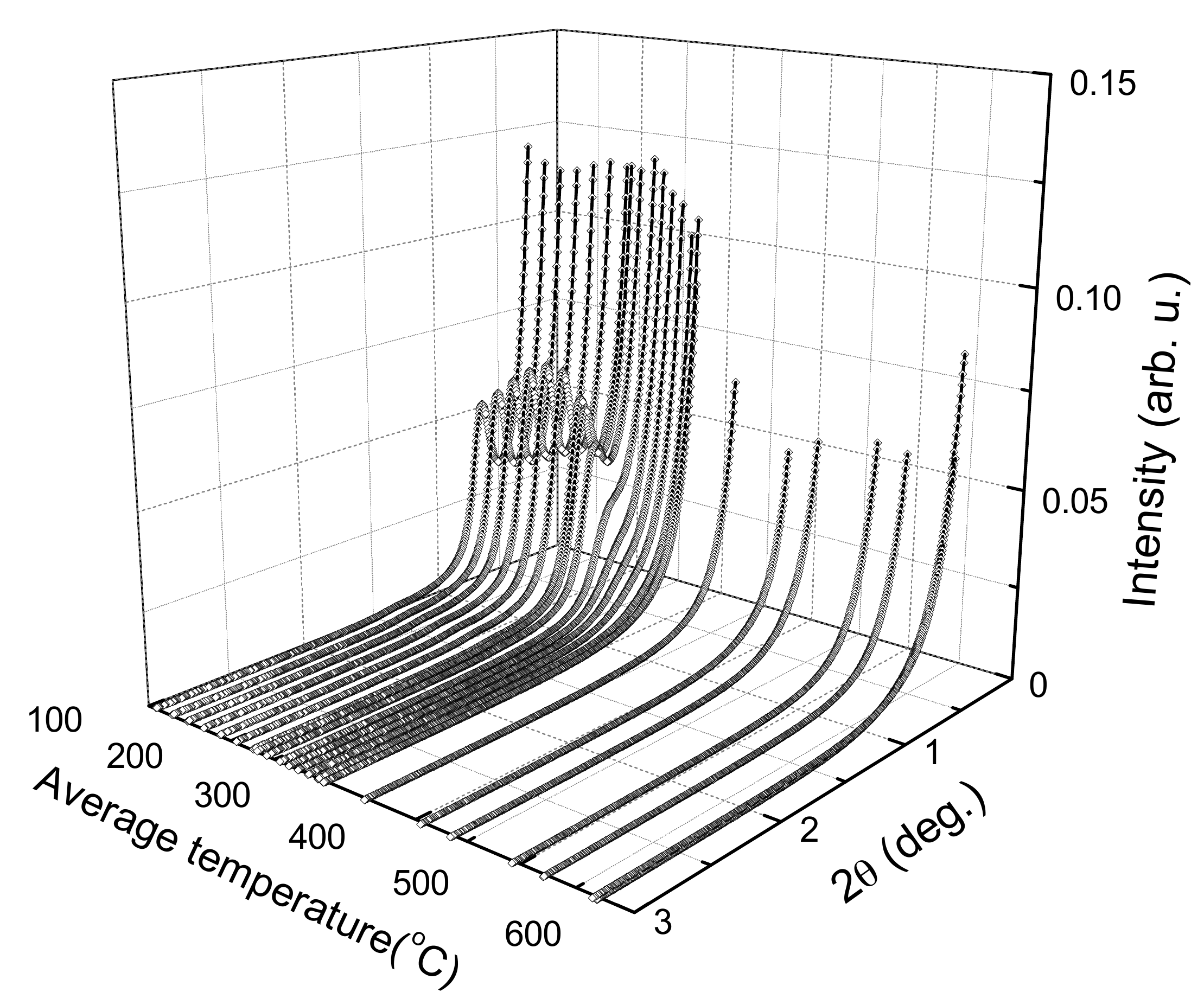}
  \caption{SAXS curves during the calcination process for the Z90AC sample, with 90\% \ce{CeO2}.}
  \label{fgr:zr90s}
\end{figure}

\begin{figure}[!ht]
\centering
  \includegraphics[width=\textwidth]{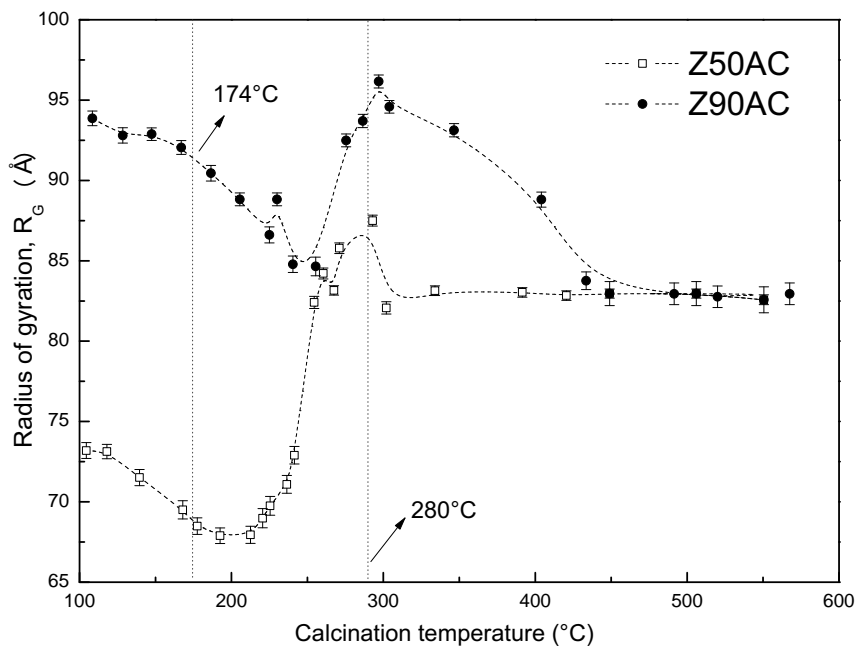}
  \caption{Radius of gyration versus calcination temperature for both samples with different \ce{CeO2} content.}
  \label{fgr:Rg}
\end{figure}

For the 90\% \ce{CeO2} content (sample Z90AC), depicted in figure \ref{fgr:zr90s}, the intensity of the peak increased as the temperature increased, reached a maximum around 170 $^{\circ}$C and then, decreased showing the loss of the ordered mesoporous structure. Figure \ref{fgr:Rg} shows the radius of gyration (R$_G$) from Guinier plot (log[I(q)] versus q$^2$) for both \ce{CeO2} contents\cite{aldo,guinier}.

The radius of gyration is related to the mean size of the scattering particles or pores in the sample\cite{aldo,guinier}. For mesoporous materials, this parameter can be related to the pore size. Before 174 $^{\circ}$C, for both \ce{CeO2} contents, there was a shrinking of the pore size, due to the decomposition of the polymer template, which is expected for \ce{SiO2} systems\cite{schuth2001}. Between 200 to 300 $^{\circ}$C the radius of gyration increased, suggesting the pore´s walls collapsed. After 300 $^{\circ}$C, both samples lost the ordered mesoporous network after the total removal of the polymer (540 $^{\circ}$C). At the end of the calcination process there was a final radius of gyration around 80 \AA for both ceria contents, showing a shrinkage of the porous structures.

\begin{figure}[!h]
\centering
  \includegraphics[width=\textwidth]{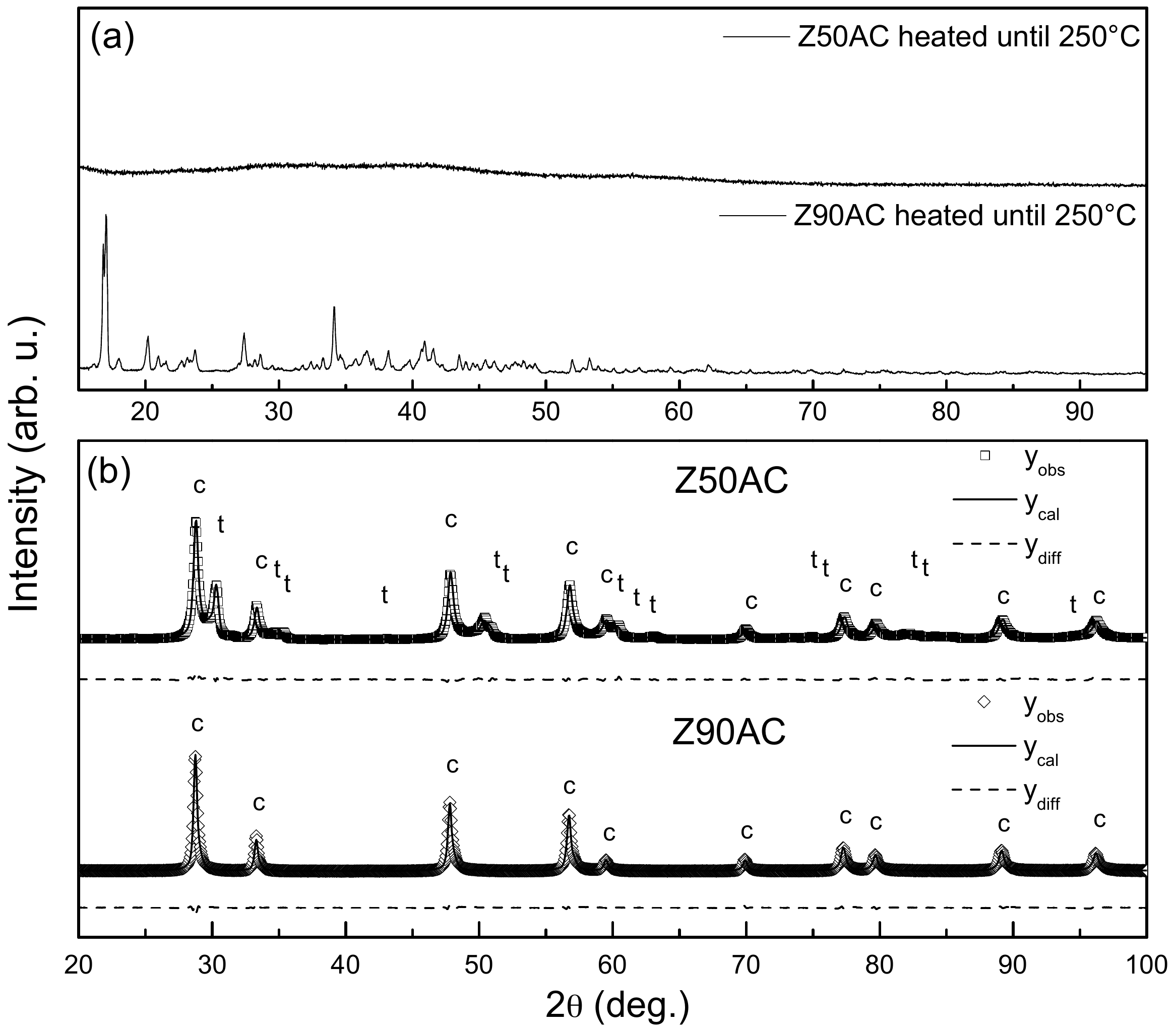}
  \caption{(a) Conventional XRD patterns for Z50AC and Z90AC heated until 250 $^{\circ}$C. (b) Synchrotron XRD pattern of calcined samples at 540 $^{\circ}$C (empty symbols) with the Rietveld fitted pattern (line) and the difference plot (dashed line), for Z50AC and Z90AC. The ``t'' denotes the tetragonal structure peaks and ``c'', the peaks for the cubic structure.}
  \label{fgr:XPDZAC}
\end{figure}

Figure \ref{fgr:XPDZAC} shows the XRD results for the Z50AC and Z90AC samples, calcined until 250 $^{\circ}$C and 540 $^{\circ}$C. 

For Z50AC sample there were no crystalline phases when it was heated until 250 $^{\circ}$C. After 540 $^{\circ}$C treatment, the crystalline structure was composed of tetragonal (P4$_2$/nmc) and cubic fluorite (Fm$\bar{3}$m) \ce{ZrO2}-\ce{CeO2} crystallographic phases\cite{duprez1999,lamas2005}.
For the Z90AC sample several peaks were observed after it was heated until 250 $^{\circ}$C, from chloride and oxychloride phases (\ce{Zr(ClO4)4} and \ce{Ce(ClO4)3}\footnote{Phases indexed with MDI Jade 6.5 software.}). There was a single cubic fluorite type phase (Fm$\bar{3}$m) after calcination until 540 $^{\circ}$C. 

Rietveld results are presented on table \ref{tbl:RAC}, showing the difference between the two ceria contents. For Z90AC there were larger lattice parameters, leading to a higher cell volume and a larger cubic phase crystallite size. From the lattice parameters it was possible to evaluate the amount of zirconia that was incorporated in to the ZDC lattice\cite{lamas2005}. For Z50AC, it was estimated that the compositions of the tetragonal and cubic phases were \ce{ZrO2}-15 mol\% \ce{CeO2} and \ce{CeO2}-5 mol\% \ce{ZrO2}, respectively. Besides, the fitting results suggested that a small part of amorphous \ce{ZrO2} was still present in this sample.

\begin{table}[!h]
\small
  \caption{Structural parameters of Rietveld analysis for 50 and 90\% \ce{CeO2} calcined samples (until 540 $^{\circ}$C). Where a and c are lattice parameters, weight percentage of crystalline phases, V is the lattice volume, D is the average crystallite size. R$_{p}$, R$_{wp}$, R$_{exp}$, $\chi^2$ and S$_{GoF}$ are the Rietveld standard agreement factors.}
  \label{tbl:RAC}
  \begin{tabular*}{0.99\textwidth}{@{\extracolsep{\fill}}llll}
    \hline
    		&\multicolumn{2}{c}{Z50AC}		&Z90AC \\
    \hline
			&Cubic		&Tetragonal	&Cubic	\\    
    Phase		&Fm$\bar{3}$m&P4$_2$/nmc	&Fm$\bar{3}$m\\
    			&73.2(4)\%	&26.84(9)\%		&100.0(6)\%\\
    \hline
    a/$\AA$		&5.4089(15)	&3.6215(5)		&5.4090(6) \\
    c/$\AA$		&-	 		&5.2160(5)		&- \\
    V/$\AA^3$	&158.236(8)	&68.41(7)		&158.364(3)\\
    D/nm		&16.3(3)	&6.63(9)		&29.8(9)\\
    \hline
    R$_{p}$		&\multicolumn{2}{c}{7.31}		&6.82\\
    R$_{wp}$	&\multicolumn{2}{c}{7.57}		&6.05\\
    R$_{exp}$	&\multicolumn{2}{c}{2.45}		&2.90\\
    $\chi^2$	&\multicolumn{2}{c}{9.56}		&5.3\\
    S$_{GoF}$	&\multicolumn{2}{c}{3.1}		&2.1\\
    \hline
  \end{tabular*}
\end{table}

\begin{figure}[!h]
\centering
  \includegraphics[width=\textwidth]{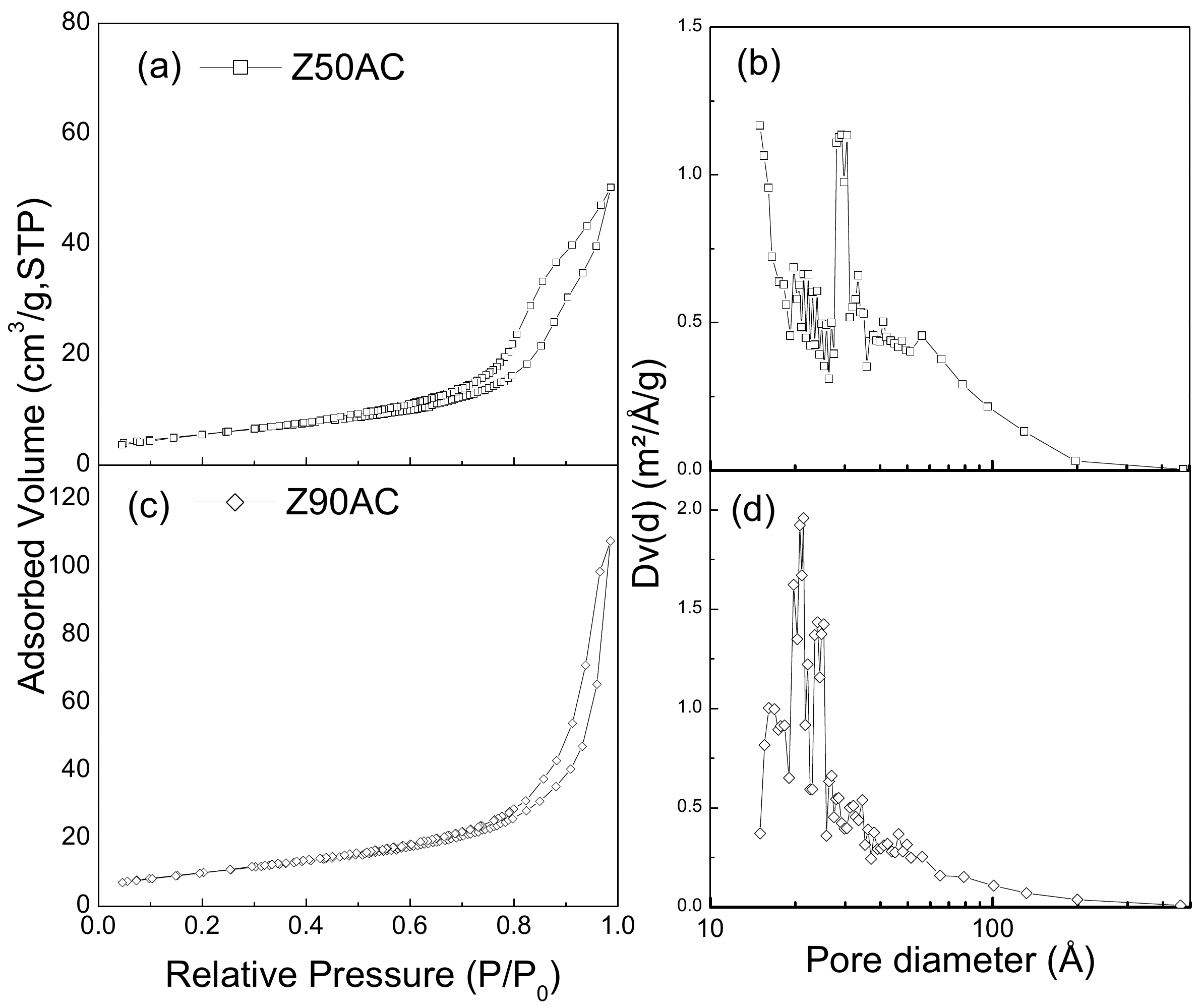}
  \caption{Nitrogen physisorption isotherm and the pore size distribution (PSD, right) calculated from the adsorption branch using the BJH method for: (a) isotherm and (b) PSD, for Z50AC sample. (c) isotherm and (d) PSD, for Z90AC sample.}
  \label{fgr:ZACNAIf}
\end{figure}

Nitrogen physisorption results are presented on figure \ref{fgr:ZACNAIf} and table \ref{tbl:ZACNAIt}. 

\begin{table}[h]
\small
  \caption{Nitrogen physisorption results: Specific surface area S$_{BET}$, pore volume V$_P$, mean pore diameter D$_P$.}
  \label{tbl:ZACNAIt}
  \begin{tabular*}{0.99\textwidth}{@{\extracolsep{\fill}}llll}
      \hline
    Samples	&S$_{BET}$	&V$_P$		&D$_P$\\
    			&(m$^2$/g)	&(cm$^3$/g)	&(\AA)\\
    \hline
    Z50AC 		&47.3 		&0.18		&28\\
    Z90AC 		&46.1 		&0.21		&21\\
      \hline
  \end{tabular*}
\end{table}

The Z50AC sample presented a type II isotherm and H$_{3}$ hysteresis. The BJH pore size distribution (PSD) showed one peak at a mean pore diameter of 28 \AA. For higher ceria content, the Z90AC sample, presented hysteresis more characteristic of H$_{1}$ type, with capillary condensation happening only at P/P$_{0}$ higher values, indicating lower pore sizes and inter-particle porosity contribution. The PSD showed a higher dispersion of pore´s size for higher ceria content, with medium pore diameter around 21 \AA. The BET specific surface area and pore volume were independent of the ceria content. The t-plot analysis showed absence of micropores for both ceria contents. The radius of gyration calculated by SAXS (fig. \ref{fgr:Rg}) was higher than the average pore diameter measured from adsorption, mostly because of inter-particle porosity contribution on the surface adsorption. 

These results indicate that the crystallization process is an important step, which guides the mesoporous structure of these oxides. And, even loosing the ordered porous structure, the ZDC samples present proper textural properties for application in catalysis and anode for SOFCs. 

\subsection{\ce{ZrO2}-\ce{CeO2}/\ce{SiO2}}

\begin{figure}[H]
\centering
  \includegraphics[width=\textwidth]{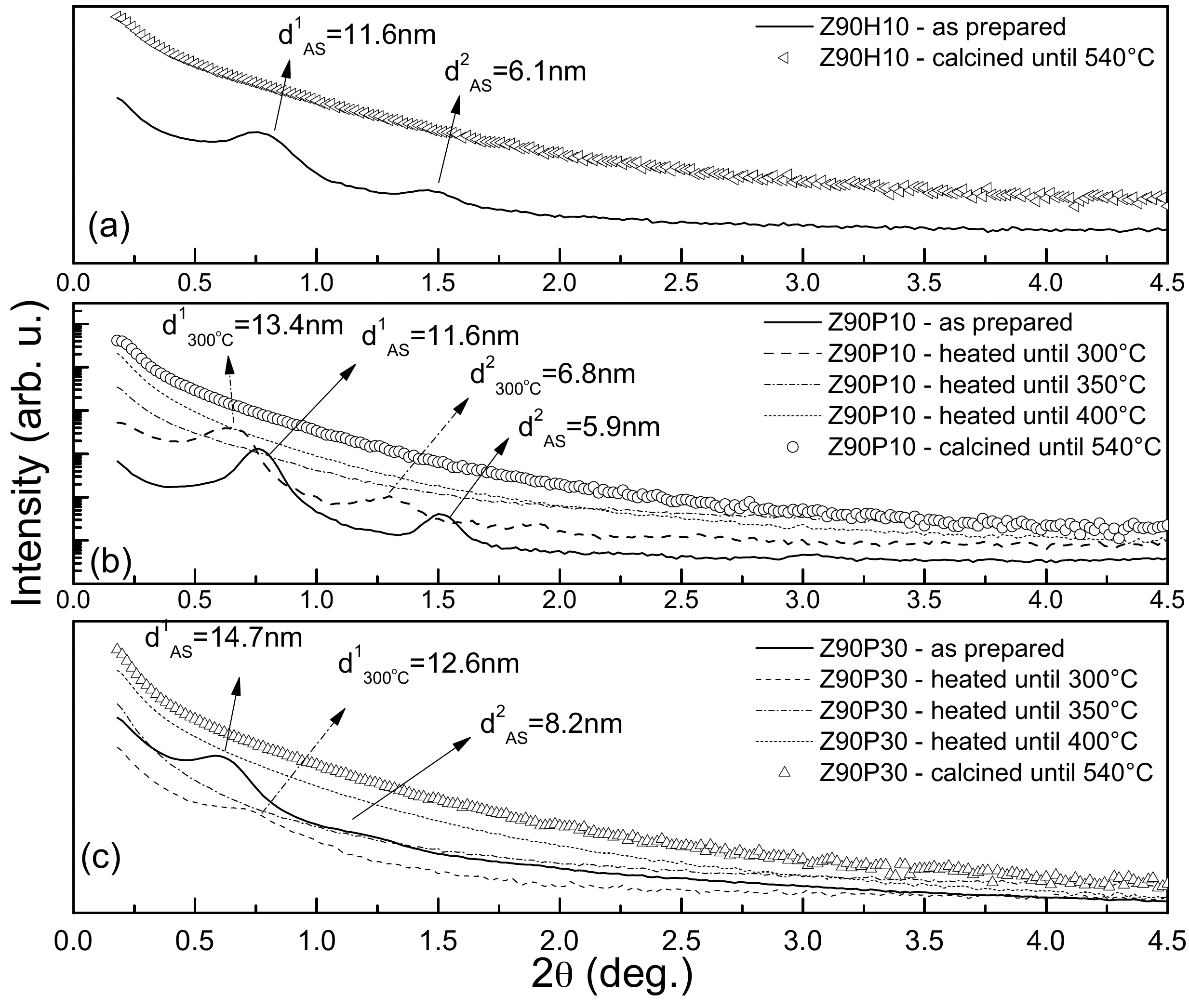}
  \caption{SAXS curves for (a) Z90H10 as prepared and calcined. Ex-situ SAXS experiment for: (a) Z90P10 and, (b) Z90P30. }
  \label{fgr:SAXS-Si}
\end{figure}

In order to increase the mechanical stability of the ZDC samples two different one-pot synthesis with low \ce{SiO2} content and 90\% \ce{CeO2} (single phase cubic structure), were tested. The SAXS results are showed in figure \ref{fgr:SAXS-Si} and table \ref{tbl:SAXS-SiT}, with results before and after calcination for all samples. 

Since the first sample from hybrid method (Z90H10) showed a disordered porous structure after calcination up to 540 $^{\circ}$C, the samples from palisade method, Z90P10 and Z90P30, were heated until intermediate temperatures (300, 350 and 400 $^{\circ}$C) in order to evaluate the ordered porous structure evolution. The presence of several peaks in the low angle/\textbf{q} region of SAXS  indicates that the material has an ordered porous structure and, more than one peak indicates which structure was formed after the synthesis. 


\begin{table}[!h]
\small
  \caption{SAXS parameters: Area under the peaks A$_{P}$, total area under the SAXS curve A$_{T}$, ordering factor f, where f = A$_{P}$/A$_{T}$, and radius of gyration, R$_G$.}
  \label{tbl:SAXS-SiT}
  \begin{tabular*}{0.99\textwidth}{@{\extracolsep{\fill}}lllll}
      \hline
    Samples	&A$_{P}$	&A$_T$	&f		&R$_G$\\
    		&(arb. u.)	&(arb. u.)	&		&($\AA$)\\
    \hline
    Z90H10 	& 			&		&	 	&\\
    as prepared	&14 	&39		&0.36 	&124\\
    calcined	&- 		&7184	&0	 	&114\\
    \hline
    Z90P10		& 		&		&	 	&\\
    as prepared	&19 	&26		&0.71	&113\\
    300$^{\circ}$C&63	&107	&0.51 	&91\\
    350$^{\circ}$C&-	&442	&0	 	&118\\
    400$^{\circ}$C&- 	&1954	&0	 	&129\\
    calcined	&- 		&5126	&0	 	&115\\
    \hline
    Z90P30		& 		&		&	 	&\\
    as prepared	&73	 	&298	&0.24	&113\\
    300$^{\circ}$C&2	&30		&0.07 	&119\\
    350$^{\circ}$C&- 	&807	&0	 	&137\\
    400$^{\circ}$C&- 	&4187	&0	 	&126\\
    calcined	&-		&9476	&0	 	&133\\
      \hline
  \end{tabular*}
\end{table}

The ordering factor f was defined as the area of the peaks after subtracting the scattering baseline (A$_{P}$) divided by the total area under the peaks without the baseline removal (A$_{T}$), so f = A$_{P}$/A$_{T}$. Since the area under the SAXS curve is proportional to the scattering pores/particles, it was possible to compare samples having higher ordering, relying on the pore structure.

Studying the interplanar distances from uncalcined samples in fig. \ref{fgr:SAXS-Si}, the results obtained for samples prepared by the hybrid and palisade methods were similar. Comparing interplanar distances d$^1$ from the first peak and d$^2$ for the second peak, it was possible to index a lamellar structure where the lattice parameter for the pores, a$_{p}$, is 1/d$_{h0}$=h/a$_{p}$. Therefore $a \sim d^1_{10}$ = 116 \AA for Z90H10 and Z90P10. The sample Z90P30 presents two peaks, closer to a 2D hexagonal structure with average lattice parameter a$_{p}$ = 180 \AA, calculated using $\frac{1}{d^2_{hk0}} = \frac{4}{3}\frac{h^2+hk+k^2}{a^2_{p}}$ (for this case, hkl is equal to 100 and 200)\cite{kresgue} . 

From hybrid and palisade method with 10 mol\% Si it was possible to infer that stirring the polymer and TEOS prior to Zr/Ce addition during preparation resulted in higher ordering factor. As the temperature increased, the total area increased, since the intensity of SAXS depends mostly on the structure density contrasts. Both samples of the palisade method maintained their ordered porous network until 350 $^{\circ}$C, where there were no more peaks on the SAXS curves. The ordering factor decreased as the temperature increased. The radius of gyration for Z90P10 decreased at 300 $^{\circ}$C, indicating a abrupt shrinkage of the structure, while for Z90P30 the radius increased at 350 $^{\circ}$C which could suggest the rupture of the walls of the 2D hexagonal structure.

\begin{figure}[H]
\centering
  \includegraphics[width=0.85\textwidth]{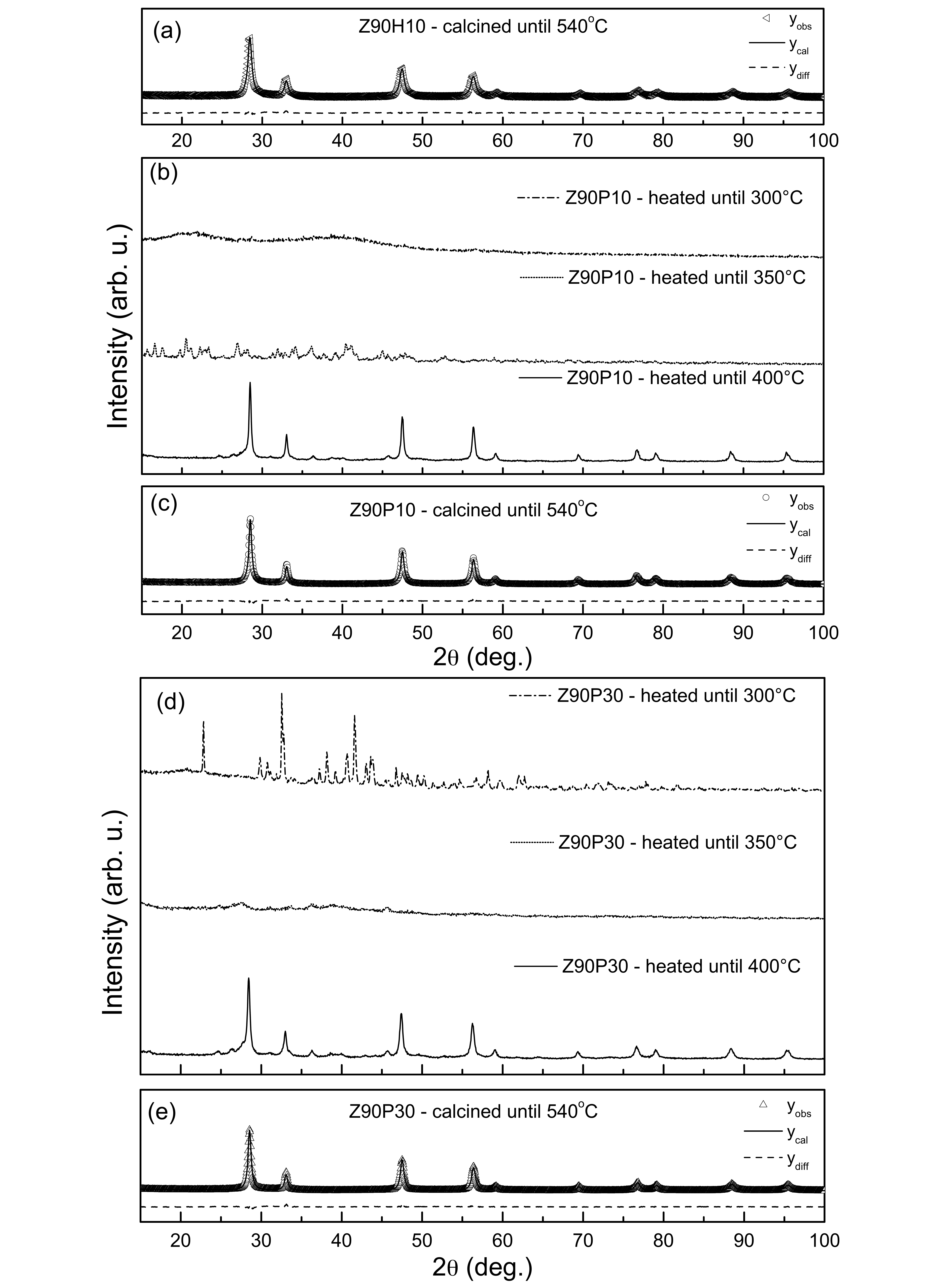}
  \caption{XRD for (a) Z90H10, calcined. Ex-situ XRD experiment for: (b) Z90P10, (c) Z90P10, calcined. Ex-situ XRD experiment for: (d) Z90P30, (e) Z90P30, calcined. Symbols represents experimental data, with the Rietveld fitted pattern (line) and the difference plot (dashed line) for calcined samples.}
  \label{fgr:XRD-Si}
\end{figure}

Figure \ref{fgr:XRD-Si} shows the XRD results for calcined Z90H10. The Z90P10 and Z90P30 samples were analysed after heat treatment up to 300, 350, 400 $^{\circ}$C and calcined (540 $^{\circ}$C). Rietveld results are presented on table \ref{tbl:RACSi}.

The XRD measurements were obtained for the same temperatures analysed by SAXS experiments, in order to study the samples' crystalline structure.

\begin{table}[!h]
\small
  \caption{Structural parameters of Rietveld analysis for 90\% \ce{CeO2} (Z90H10, Z90P10 and Z90P30) calcinated samples (until 540 $^{\circ}$C). Where a and c are lattice parameters, weight percentage of crystalline phases, V is the lattice volume, D is the average crystallite size. R$_{p}$, R$_{wp}$, R$_{exp}$, $\chi^2$ and S$_{GoF}$ are the Rietveld standard agreement factors.}
  \label{tbl:RACSi}
  \begin{tabular*}{0.99\textwidth}{@{\extracolsep{\fill}}llll}
    \hline
    			&Z90H10		&Z90P10		&Z90P30 \\
    \hline
				&Cubic		&Cubic		&Cubic\\    
    Phase		&Fm$\bar{3}$m&Fm$\bar{3}$m&Fm$\bar{3}$m\\
    			&100.0(7)\%	&100.0(5)\%	&100.0(8)\%\\
    \hline
    a/$\AA$		&5.4150(14)	&5.4089(11)	&5.4094(11)\\ 
    V/$\AA^3$	&158.38(7)	&158.24(6)	&158.29(6)\\
    D/nm		&18.59(9)	&29.5(4)	&27.7(3)\\
    \hline
    R$_{p}$		&8.24		&9.71		&9.19\\
    R$_{wp}$	&10.5		&12.0		&11.9 \\
    R$_{exp}$	&5.15		&5.74		&5.85 \\    
    $\chi^2$	&4.2		&4.4		&4.1\\
    S$_{GoF}$	&1.9		&2.1		&2.0\\
    \hline
  \end{tabular*}
\end{table}

For Z90P10 the crystallization started only at 350 $^{\circ}$C, later than previous sample with the same ceria content. Z90P30 at 350 $^{\circ}$C presented an amorphous pattern, showing a even later crystallization, at 400 $^{\circ}$C. Sample Z90H10 showed a smaller crystallite size. From Rietveld results there were no significant changes on the overall crystallographic structure of the ZDC synthesized with Si, the lattice parameters are similar from cubic single-phase solid solutions synthesized by gel-combustion routes\cite{lamas2005}. 

\begin{figure}[!ht]
\centering
  \includegraphics[width=0.8\textwidth]{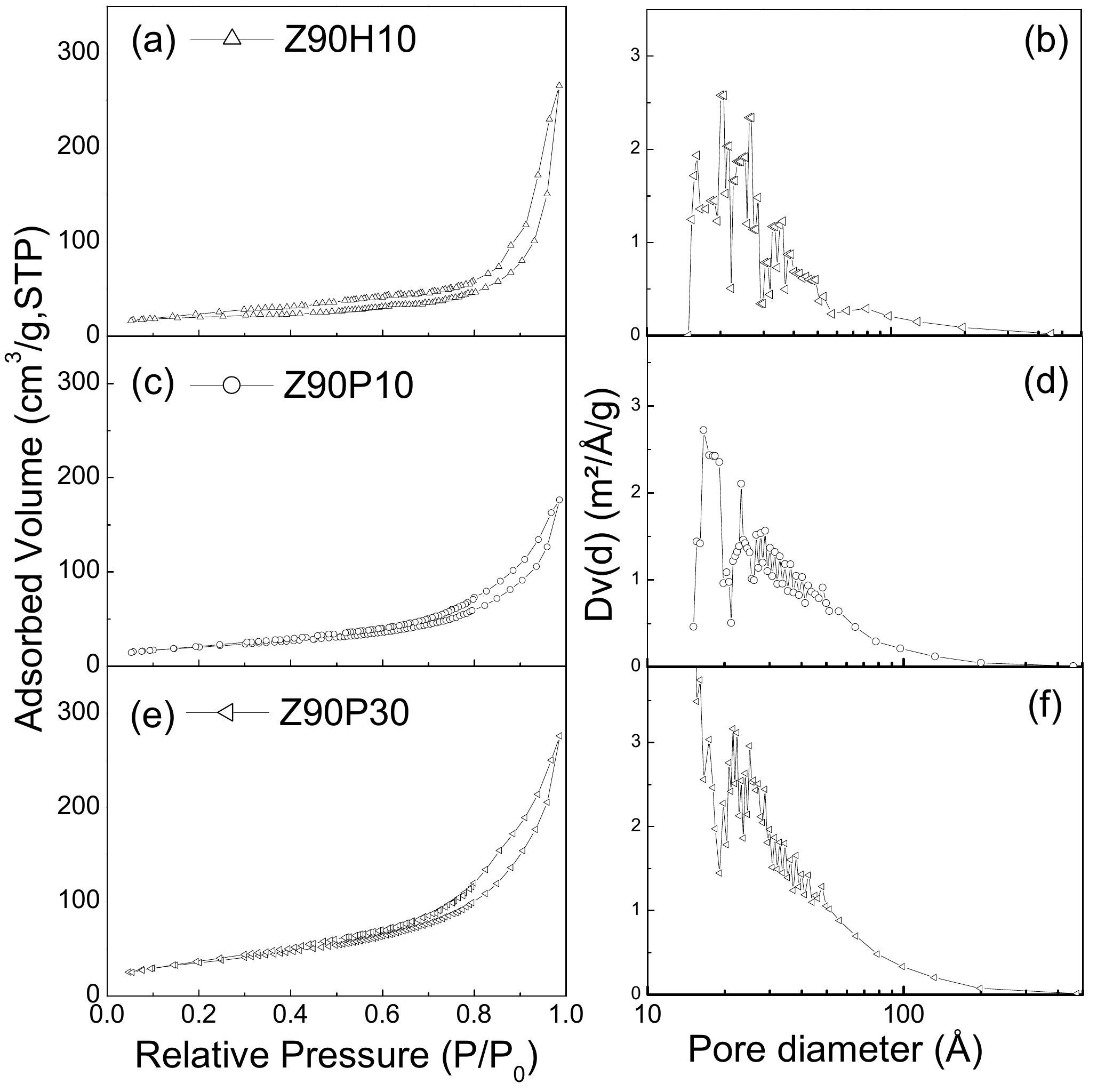}
  \caption{Nitrogen physisorption isotherms and the PSD for: Z90H10 (a) isotherm, and (b) PSD.  Z90P10 (c) isotherm, and (d) PSD. Z90P30 (e) isotherm, and (f) PSD.}
  \label{fgr:NAI-Si}
\end{figure}

N$_2$ physisorption results are presented on figure \ref{fgr:NAI-Si} and table \ref{tbl:NAI-SiT}.

Hybrid method sample resulted in a type H$_{4}$ hysteresis, while both palisade samples presented hysteresis loops that could be identified as H$_{2}$ and H$_{3}$ types. From PSD curves a higher silica content resulted a narrower pore size distribution.

\begin{table}[!h]
\small
  \caption{Nitrogen adsorption results: Specific surface area S$_{BET}$, pore volume V$_P$, mean pore diameter D$_P$.}
  \label{tbl:NAI-SiT}
  \begin{tabular*}{0.99\textwidth}{@{\extracolsep{\fill}}llll}
      \hline
    Samples	&S$_{BET}$	&V$_P$		&D$_P$\\
    		&(m$^2$/g)	&(cm$^3$/g)	&(\AA)\\
    \hline
    Z90H10 	&67.5 		&0.43		&23\\
    Z90P10	&72.7 		&0.27		&26\\
    Z90P30	&128.2 		&0.41		&24\\

      \hline
  \end{tabular*}
\end{table}

From t-plot analysis\cite{tplot}, micropores were detected in the Z90H10 sample (V$^{t-plot}_P$ = 0.012 cm$^3$/g and S$^{t-plot}$ = 8.1 m$^2$/g), while Z90P10 and Z90P30 only presented mesopores. The addition of the Zr/Ce precursors after the previous P-123 and TEOS solution resulted in a higher surface area, but slightly lower pore volume. Higher silica content resulted in higher surface area and pore volume, showing the advantage of a silica palisade to improve the material textural properties.

\begin{figure}[!ht]
\centering
  \includegraphics[width=\columnwidth]{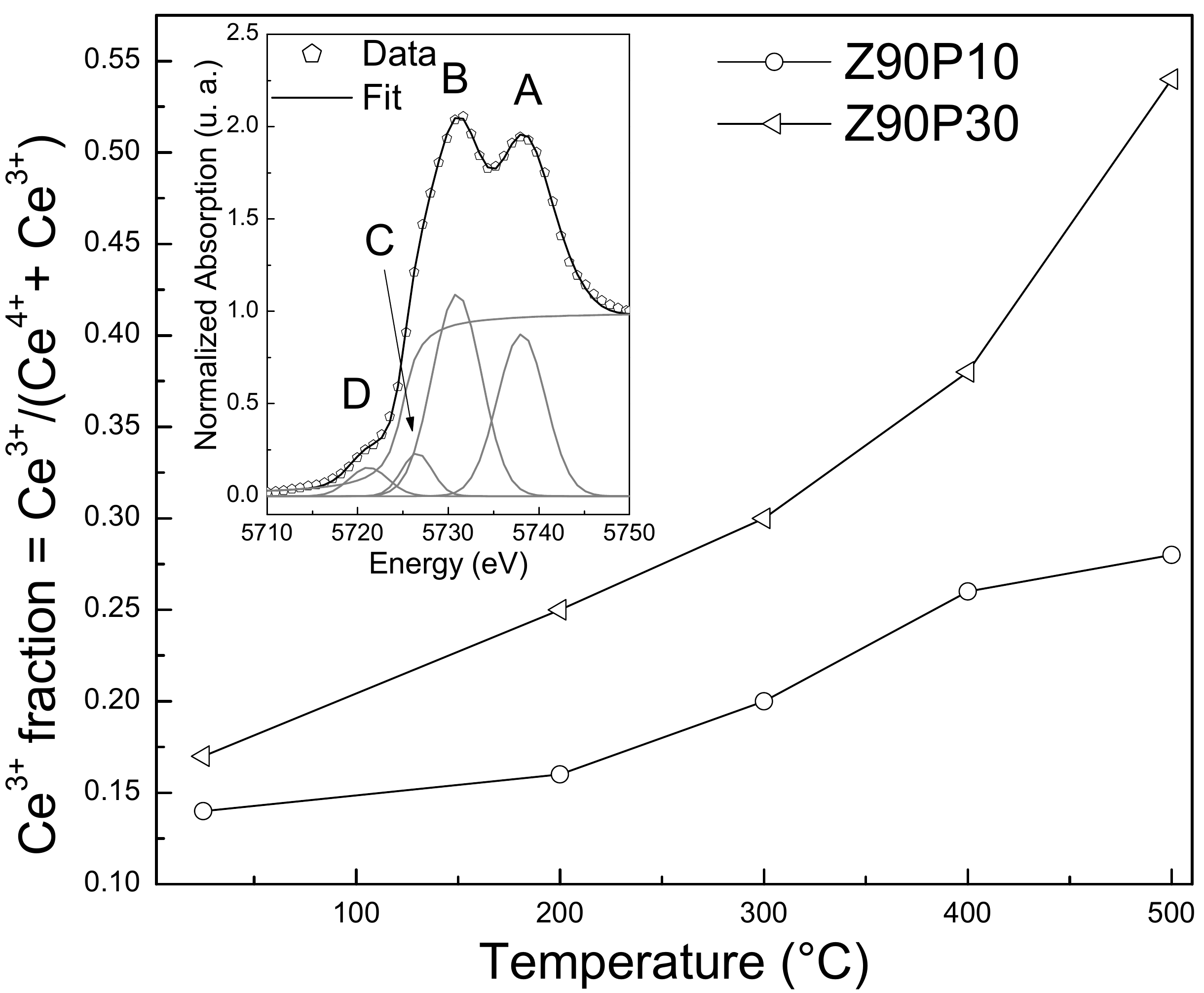}
  \caption{Ce$^{3+}$ fraction for Z90P10 and Z90P30 samples as a function of temperature, under temperature programmed reduction (TPR) reaction with 5\% \ce{H2}/He. Inset shows the \ce{CeO2} XANES spectrum at Ce L$_{III}$-edge.}
  \label{fgr:XANES}
\end{figure}

In situ XANES results are presented on figure \ref{fgr:XANES}. The analysis method described by Zhang et. al\cite{zhang2004} was used to quantify the fraction of Ce$^{3+}$ in the samples. It consists of a least-squares fit of the experimental data with the WinXAS software, using four Gaussian profiles and one arctangent function (fig. \ref{fgr:XANES}, inset)\cite{winxas}.

Pure \ce{CeO2} spectrum at L$_{III}$-edge has two main peaks, A and B, which are related to Ce$^{4+}$ final states from 2p to 5d transitions. The peak C occurs due to dipole transitions and it is characteristic of the Ce trivalent state. The pre-edge peak D corresponds to multiple scattering processes\cite{zhang2004,yoshida2004,shanin2005,acuna2010}. Therefore, the Ce$^{3+}$ fraction is equal to the relation of the peak areas, such as  C/(A+B+C), where A, B and C represent the area of the fitted peaks. Fig. \ref{fgr:XANES} shows that with the increase of temperature under reduction conditions there is a shift from Ce$^{4+}$ to Ce$^{3+}$ for both samples. At 500 $^{\circ}$C the sample Z90P10 reached a reduction percentage of 28\%, and sample Z90P30, 54\%.

Although there is not total reduction of Ce$^{3+}$ to Ce$^{4+}$ in the studied temperature range, the increase of the silica content from 10 to 30 mol$\%$ and the improvement on the superficial
area (from from 73 to 120 m$^2$/g, respectively) promoted a higher reduction of Ce$^{4+}$ under lower temperatures compared to materials synthesized via gel-combustion routes\cite{acuna2010,geno2012}.

\section{Conclusions}

In this work new syntheses of ZDC were reported, using soft template methods. Three new syntheses methods with low \ce{SiO2} content were developed to increase mechanical stability of the ZDC porous structure. The lack of an ordered structure of the pores was correlated to the amorphous-crystalline transition of \ce{ZrO2}-\ce{CeO2} at 300 $^{\circ}$C. Although there were no ordered pores, the samples showed proper textural properties for applications in catalysis, as well as SOFC anodes. A higher pore ordering was obtained and kept up to 350 $^{\circ}$C, compared to other materials reported in the literature. Appropriate textural properties were obtained, specially higher specific surface area and pore volume compared to sol-gel ZDC samples\cite{geno2012}. And finally, in situ XANES results revealed a high reduction rate of Ce$^{3+}$ to Ce$^{4+}$ at 500 $^{\circ}$C under TPR, which is directly correlated to higher catalytical activity.

\paragraph{Acknowledgments} Thanks to CNPq and FAPESP for supporting this research. Thanks to Dr. R. O. Fuentes and Dr. L. M. Acu\~na for synchrotron data collection. The LNLS is acknowledged for the use of D11A-SAXS1 (project number 7124), D10B-XPD (project number 2960) and D04B-XAFS1 (project number 9218) beamlines. 

\section*{References}

\bibliography{zdcsi}

\end{document}